\documentclass[useAMS,usenatbib,usegraphicx,uselongtable]{mn2e}

\title[Erratum]{Erratum: Herschel SPIRE observations of the TWA brown dwarf disc 2MASSW J1207334-393254}%\thanks{{\it Herschel} is an ESA space observatory with science instruments provided by European-led Principal Investigator consortia and with important participation from NASA.}}
\author[Riaz et al.]
{B. Riaz,$^{1}$ G. Lodato,$^{2}$ D. Stamatellos,$^{3}$ J. E. Gizis$^{4}$  \\
$^{1}$Centre for Astrophysics Research, Science \& Technology Research Institute, University of Hertfordshire, Hatfield, AL10 9AB, UK \\
$^{2}$Dipartimento di Fisica, Universita Degli Studi di Milano, Via Celoria, 16, Milano, 20133, Italy \\
$^{3}$School of Physics \& Astronomy, Cardiff University, Cardiff, CF24 3AA, UK \\
$^{4}$Department of Physics and Astronomy, University of Delaware, Newark, DE 19716, USA \\
 }

\begin{document}

\date{}

\pagerange{\pageref{firstpage}--\pageref{lastpage}} \pubyear{2012}

\maketitle

\label{firstpage}

\begin{abstract}

We have revised our analysis of the SPIRE observations of 2MASSW J1207334-393254 (2M1207). Recent PACS observations show a bright source located $\sim$25$\arcsec$ east of 2M1207. There are issues in terms of the detection/non-detection of the bright source when comparing the {\it Spitzer}, WISE, and PACS observations. It is apparently inconsistent, perhaps due to variability or low signal-to-noise of the data. We have looked into the possible misidentification of the target, and have revised the measured SPIRE fluxes and the disc parameters for 2M1207. We have also reviewed which among the various formation mechanisms of this system would still be valid. 
%We have revised the measured SPIRE fluxes and the disc parameters for 2M1207. We have also reviewed which among the various formation mechanisms of this system would still be valid. 

%However, the bright source appears variable in terms of detection/non-detection when comparing the {\it Spitzer}, WISE, and PACS observations. 

%The separation between 2M1207 and the bright object centroid is comparable to the offset between the nominal and actual position of the target, and the target may have been misidentified. We have revised the measured SPIRE fluxes and the disc parameters for 2M1207, and have also reviewed which among the various formation mechanisms of this system would still be valid. 

\end{abstract}


\noindent {\bf Observations and Data Analysis}
\vspace{0.05in}
%\section{Observations and Data Analysis}

\begin{figure}
\includegraphics[width=53mm]{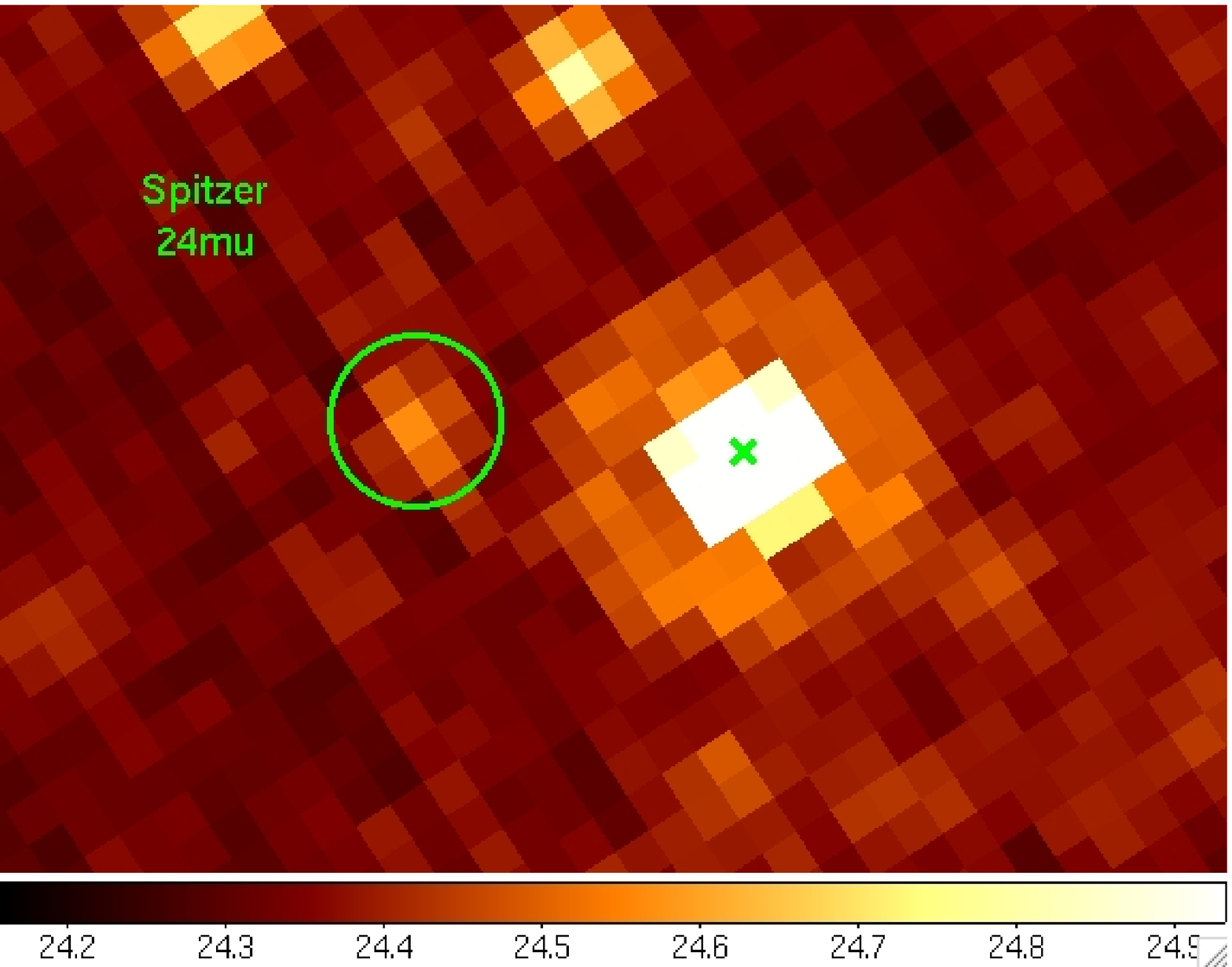} \\
\includegraphics[width=53mm]{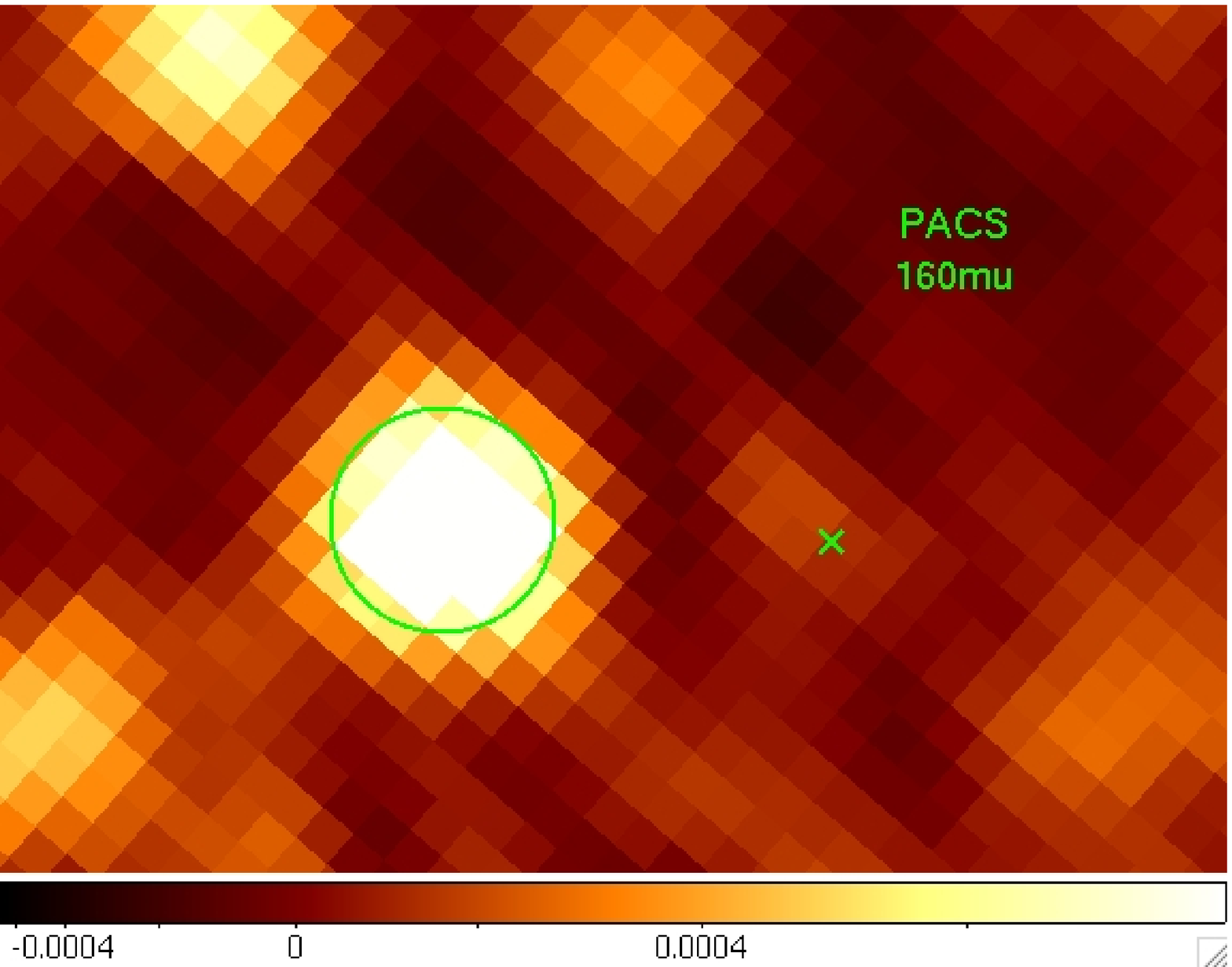} \\
\includegraphics[width=53mm]{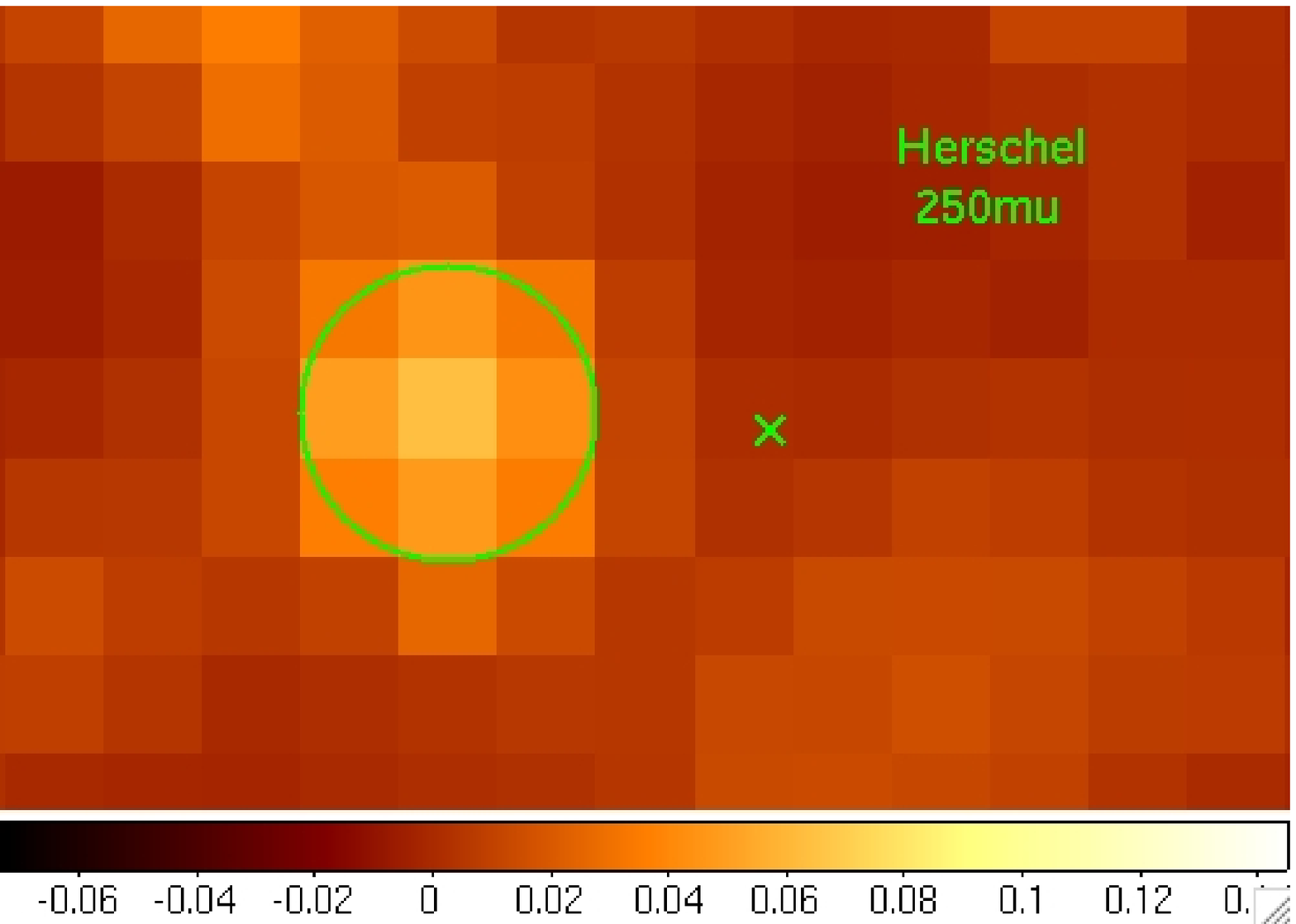} \\
\includegraphics[width=53mm]{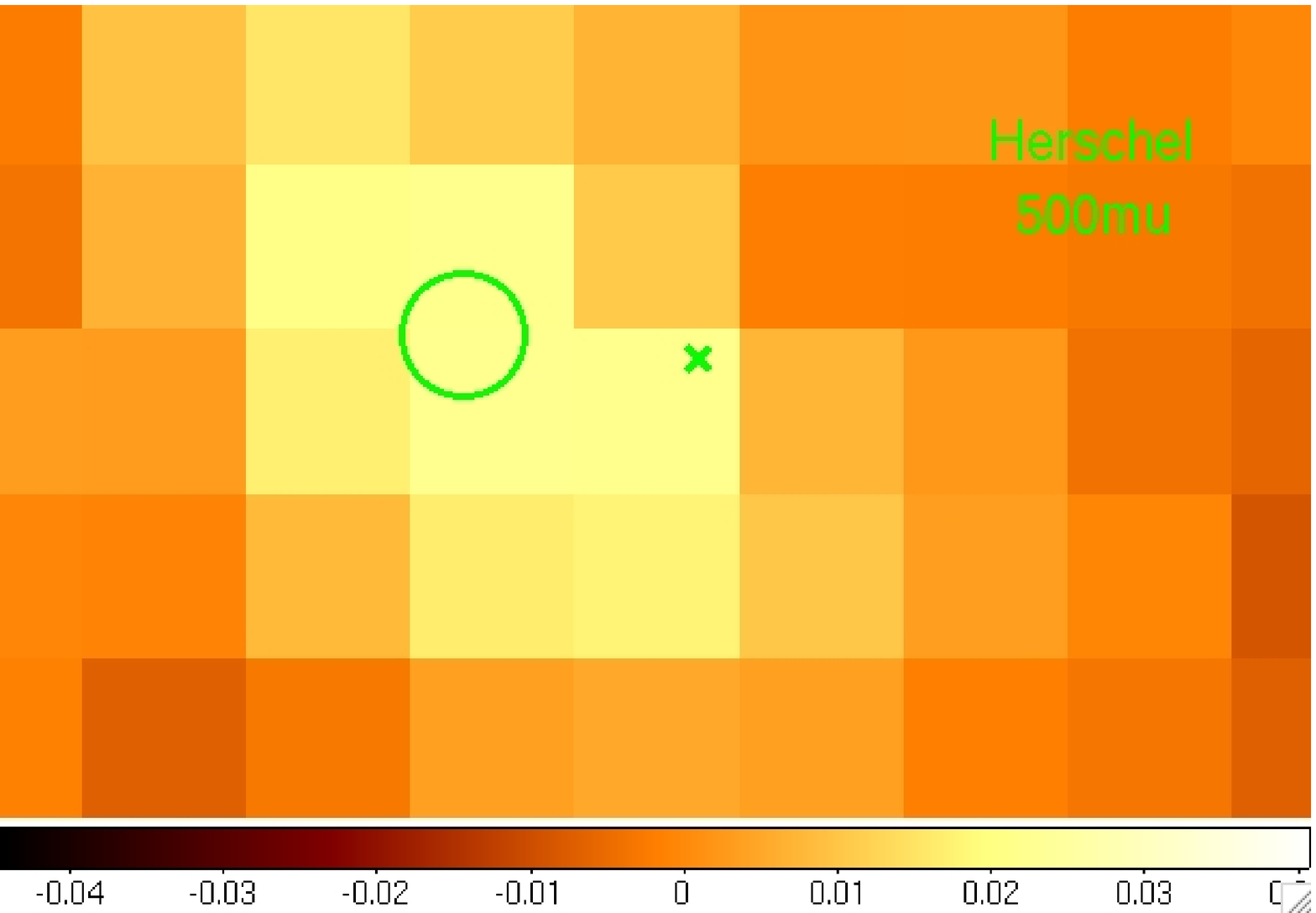} \\
    \caption{2M1207 images: ({\it top}) {\it Spitzer} 24$\micron$; ({\it second}) PACS 160$\micron$; ({\it third}) SPIRE 250$\micron$; ({\it bottom}) SPIRE 500$\micron$. 2M1207 is marked by a cross, Source120735 is marked by a circle. The pixel scale is 2.4$\arcsec$ pix$^{-1}$ in the {\it Spitzer} 24$\micron$ band, 0.4$\arcsec$ pix$^{-1}$ in PACS, 6$\arcsec$ pix$^{-1}$ in the SPIRE 250$\micron$ band, and 14$\arcsec$ pix$^{-1}$ in the 500$\micron$ band. In all images, North is up and East is to the left.  }
   \label{images}
 \end{figure}

\noindent We had reported in Riaz et al. (2012) a detection for 2MASSW J1207334-393254 (2M1207) in the {\it Herschel} SPIRE bands of 250 and 350$\micron$. A parallel study conducted by Harvey et al. (2012) based on PACS 70 and 160$\micron$ observations shows a bright source at RA = 12:07:35.183; Dec = -39:32:52.64 (120735), located $\sim$25$\arcsec$ east of 2M1207. This is an unclassified source with no SIMBAD matches (other than 2M1207) within 30$\arcsec$. There is no detection for this object in the 2MASS bands. It is detected ($>$2-$\sigma$) in the {\it Spitzer} IRAC 3.6 and 4.5$\micron$ bands, but is undetected in the 5.8 and 8$\micron$ bands. We have also checked the WISE images and there is a detection ($>$2-$\sigma$) for this source at 3.4 and 4.6$\micron$, but it is undetected in the 12 and 22$\micron$ bands. It was faintly detected at a 1-$\sigma$ level in the {\it Spitzer} 24$\micron$ image, but was undetected in the {\it Spitzer} MIPS 70 and 160$\micron$ bands. The {\it Spitzer} observations were obtained in 2007, whereas the PACS 70 and 160$\micron$ observations were taken in 2010. The object 120735 is an $\sim$8mJy source in the PACS 70$\micron$ observation, whereas the 1-$\sigma$ confusion noise in the {\it Spitzer} 70$\micron$ image is $\sim$2mJy. Therefore a 4-$\sigma$ detection with {\it Spitzer} would have been possible. The non-detection of this source in the {\it Spitzer} MIPS data could be due to possible variability or the low signal-to-noise of the data. We do not know the nature of this source. If it is a galaxy then it is unlikely to be variable, but then the non-detection in some of the bands is puzzling. To check the field in the SPIRE images for 2M1207, we had used the previously available {\it Spitzer} MIPS observations, since the PACS data was not public at the time the analysis was done. 2M1207 is a prominently bright detection in the 24$\micron$ band, while the source 120735 is a faint detection at a 1-$\sigma$ level (Fig.~\ref{images}). Comparing the {\it Spitzer} MIPS and the SPIRE fields, we found no clear source detection at the nominal position for 2M1207, while a bright detection was seen at the location of the source 120735. The separation between 2M1207 and the bright object centroid is comparable to the offset from the nominal position of the target ($\sim$14.3$\arcsec$ or $\sim$2.4 pixel). The offset is also identical in all three SPIRE bands. Considering the marginal (1-$\sigma$) detection of the source 120735 in the {\it Spitzer} 24$\micron$ band, its non-detection at {\it Spitzer} 70$\micron$, and the mentioned offset in the SPIRE bands, we had identified the bright object in the SPIRE images as 2M1207. A comparison now with the PACS images indicates this to be a misidentification. There is the possibility of variability for the source 120735 and the contamination from it cannot be properly accounted for, but considering how bright the object 120735 is in the PACS images, its emission is likely to dominate the SPIRE photometry. Given these uncertainties, we have revised the SPIRE fluxes for 2M1207 by measuring the emission at its nominal position, without considering the positional offset. All SPIRE measurements are upper limits (Table~\ref{fluxes}). The 500$\micron$ upper limit is the same as estimated in the original paper. Both objects lie in a confusion noise dominated region in the 500$\micron$ image (Fig~\ref{images}; bottom panel), and the flux value at the nominal position of 2M1207 and the object 120735 location is the same. The flux value at 500$\micron$ is higher than 250 or 350$\micron$ because of the higher confusion noise in this band. 

%, and the positional offset for 2M1207 in the SPIRE bands
%, which is the typical {\it Herschel} pointing uncertainty

%There is no detection for this source in the 2MASS bands. 

%We do not know the nature of this source. If it is a galaxy then it is unlikely to be variable, but then the non-detection in some of the bands is puzzling. 

%We had checked the {\it Herschel} data for some of the other targets we have, and the offset from the nominal positions is typically $\sim$1$\arcsec$--2$\arcsec$. 

%However, the offset between the nominal source location for 2M1207 and the bright object centroid is comparable to the difference in the actual ra, dec and the raNominal, decNominal in the metadata ($\sim$14.3$\arcsec$ or $\sim$2.4 pixel). 

%There may also be the possibility that the two objects overlap in the SPIRE bands, but considering how bright Source120735 is in the PACS images, its emission will likely dominate the SPIRE composite photometry in such a scenario. 

%Source120735 appears as a variable object when comparing the {\it Spitzer}, WISE, and PACS observations.

%It could be a transition object where emission at the short wavelengths could be from the starlight and the longest wavelengths from cold dust, while the intermediate depression could be due to an inner opacity hole.  

 %The size of the bar in the bottom panel is the same as the position offset in this image. 

%\section{Disk Modeling}

%\section{Revision to Section \S3}

%\clearpage

\vspace{0.05in}
\noindent {\bf Revision to Section \S 3}
\vspace{0.01in}

\begin{figure}
\includegraphics[width=60mm]{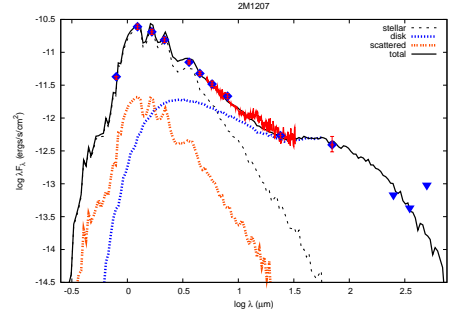} 
    \caption{The best model fit for 2M1207A disc (black line). Also shown is the contribution from the disc (blue) and the stellar photosphere (grey). The Spitzer/IRS spectrum is shown in red. The optical, near- and mid-infrared photometry plotted is listed in Riaz \& Gizis (2007). }
   \label{model}
 \end{figure}

\noindent Using the sub-mm upper limits, the best model fit is for an outer disc radius of 50 AU and a disc mass of 0.1 $M_{Jup}$ (Fig.~\ref{model}). The present model fit is the same as presented in Riaz \& Gizis (2008). We refer the readers to that paper for details on the rest of the fitting parameters. 

%(based on the lowest $\chi^{2}$ value) 

%A 100 AU model for a mass of 0.5 $M_{Jup}$ is also a good fit, though at a slightly higher $\chi^{2}$ value than the best fit. The 10 and 30 AU models are not a good fit to the longer part of the IRS spectrum for wavelengths $>$10$\micron$. The optical depth is too large for these models even if the disc mass is reduced to 0.01 $M_{Jup}$, and misses the 70-350$\micron$ points. 

%Increasing the disc mass for the same radius increases the optical depth beyond $\sim$30$\micron$, while increasing the radius for the same disc mass reduces the optical depth longward of $\sim$30$\micron$.

\begin{table}
%\begin{minipage}{30cm}
\caption{Observations for 2M1207}
\label{fluxes}
\begin{tabular}{cc}
\hline
Band & Flux [mJy]  \\ \hline
%{\it I} & 1.13$\pm$0.1 \\
%{\it J} &  10.10$\pm$0.89 \\
%{\it H} &  11.35$\pm$1.0 \\
%{\it K} & 11.12$\pm$0.98 \\
%3.6$\mu$m &  8.49$\pm$0.32 \\
%4.5$\mu$m & 7.15$\pm$0.26 \\
%5.8 $\mu$m  &  6.36$\pm$0.06 \\
%8$\mu$m  &  5.74$\pm$0.21 \\
%24$\mu$m  &  4.32$\pm$0.03 \\ 
%70$\mu$m & 9$\pm$4 \\
250$\micron$ & $<$5.2  \\
350$\micron$ & $<$5  \\
500$\micron$ & $<$16  \\
\hline
\end{tabular}
%\end{minipage}
\end{table}

%\begin{table}
%\caption{Disc parameters}
%\label{results}
%\begin{tabular}{ccc}
%\hline

%Parameter  & Value \\ \hline
%$\beta$ & 1.1$\pm$0.01 \\
%$h_{0}$ & 0.1$\pm$0.1 \\
%$R_{min}$ & 1 $R_{sub}$ ($\sim$3 $R_{*}$) \\
%$R_{max}$ & 50 AU \\
%$M_{disc}$ & 0.1 -- 0.5 $M_{Jup}$ \\
%{\it i} & 57$\degr$ -- 69$\degr$ \\
%\hline
%\end{tabular}
%\end{table}

%\section{Discussion}

\vspace{0.05in}
\noindent {\bf Revision to Section \S 4.1}
\vspace{0.05in}

\noindent A disc mass of $\sim$0.1 $M_{Jup}$ places 2M1207 among the weaker discs in Taurus. The relative disc mass for 2M1207 [{\it log} ($M_{disc}/M_{*}$) = -2.4] is comparable to the weakly accreting systems in TWA, such as, Hen 3-600. 

%A disc mass of $\sim$0.1 $M_{Jup}$ places 2M1207 among the weaker discs in Taurus, thus suggesting that the disc mass may be dependent on the age of the system. Though as noted previously, a confirmation of these trends requires larger samples of brown dwarfs across a wide range in ages. 

\vspace{0.05in}
\noindent {\bf Revision to Section \S 4.2}
\vspace{0.05in}

\noindent The core accretion mechanism is still unlikely for the same reasons as discussed in Section \S 4.2.1 in the original version of the paper (Riaz et al. 2012). For disc fragmentation (Section \S 4.2.2), our argument was that even a very low mass disc could produce a fragment of $\sim$0.035$M_{Jup}$, which can then grow over time to form a 5 $M_{Jup}$ mass object. The main requirement for such a case is for the initial mass of the disc to be higher than its current estimate (at least 10--20 $M_{Jup}$). An upper limit on the disc mass of $\sim$0.1 $M_{Jup}$ thus still does not rule out disc fragmentation, since the system is relatively old and we have not observed it during its early stages when fragmentation could have occurred ($<$0.1 Myr). The alternative mechanisms, as discussed in Section \S 4.2.3 in the original paper, would still be applicable. %These mechanisms are related to the formation of the two components of the 2M1207 system independently, and being tidally locked together after formation. {\bf However, the applicability of the star-disc model will be questionable now, considering that a much lower disc mass would result in decreasing the kinetic energy of the disc to capture another star and form a bound system.}

%It is difficult to relate current conditions to the conditions at the time of formation. 

\vspace{0.05in}
\noindent {\bf Revision to Section \S 4.3}
\vspace{0.05in}

%The results from disc modeling again indicate that small outer radii of 10-30 AU do not provide a good fit to the data. This could weaken the possibility of 2M1207A to have formed via the ejection mechanism, though recent simulation results indicate that a clear distinction cannot be made between a star-like and an ejection formation mechanism for brown dwarfs, based on the information on the spatial extent of the disc (e.g., Bate 2009). 

\noindent We had estimated in Riaz et al. (2012) an outer disc radius of 50 -- 100 AU. Our current model fit for $R_{max}$ of 50 AU is consistent with the previous estimate, and thus the possibility of the planetary mass companion truncating the disc is still applicable. We refer the readers to the discussion in Section \S 4.3 in the original version of the paper. 

%\vspace{0.05in}
%\noindent {\bf Summary}

%\noindent We have revised the SPIRE fluxes for 2M1207 due to a misidentification of the target. We estimate an upper limit to the disc mass of 0.1 $M_{Jup}$. The applicability of the formation mechanisms discussed in Section \S 4 in the original paper has been reviewed. 

\label{lastpage}


\begin{thebibliography}{}

%\bibitem[\protect\citeauthoryear{Bate}{2009}]{bate} Bate, M. 2009, MNRAS, 392, 590
\bibitem[\protect\citeauthoryear{Harvey et al.}{2012}]{harvey} Harvey, P. et al. 2012, ApJL, 744, 1
\bibitem[\protect\citeauthoryear{Riaz \& Gizis}{2007}]{rg07} Riaz, B. \& Gizis, J. 2007, ApJ, 661, 354
\bibitem[\protect\citeauthoryear{Riaz \& Gizis}{2008}]{rg08} Riaz, B. \& Gizis, J. 2008, ApJ, 681, 1584
\bibitem[\protect\citeauthoryear{Riaz et al.}{2012}]{r12} Riaz, B.; Lodato, G.; Stamatellos, D.; Gizis, J. E., 2012, MNRAS, 422, L6

\end{thebibliography}
\end{document}